\documentclass[a4paper,fleqn]{cas-dc}

\usepackage[numbers]{natbib}
\usepackage{siunitx}
\def\tsc#1{\csdef{#1}{\textsc{\lowercase{#1}}\xspace}}
\tsc{WGM}
\tsc{QE}
\tsc{EP}
\tsc{PMS}
\tsc{BEC}
\tsc{DE}
\begin{document}
\let\WriteBookmarks\relax
\def\floatpagepagefraction{1}
\def\textpagefraction{.001}

\ExplSyntaxOn
\cs_gset:Npn \__first_footerline:
  { \group_begin: \small \sffamily \__short_authors: \group_end: }
\ExplSyntaxOff 

\shorttitle{Thermal gradient effect on hydrogen transport in tungsten}

\shortauthors{S Alturk et~al.}

\title [mode = title]{Thermal gradient effect on hydrogen transport in tungsten} 
%\tnotemark[1,2]

%\tnotetext[1]{This document is the results of the research
%   project funded by the National Science Foundation.}

%\tnotetext[2]{The second title footnote which is a longer text matter
%   to fill through the whole text width and overflow into
%   another line in the footnotes area of the first page.}

\author[1]{Sanad Alturk}
%[type=editor,
%                        auid=000,bioid=1,
%                        prefix=Sir,
%                        role=Researcher,
%                        orcid=0000-0001-7511-2910]

%\cormark[1]


%\ead{cvr_1@tug.org.in}

%\ead[url]{www.cvr.cc, cvr@sayahna.org}

\credit{Conceptualization of this study, Methodology, Software}

\affiliation[1]{School of Mechanical and Automotive Engineering, Clemson University, Clemson, SC, 29362, USA}
%{organization={Elsevier B.V.},
%    addressline={Radarweg 29}, 
%    city={Amsterdam},
    % citysep={}, % Uncomment if no comma needed between city and postcode
%    postcode={1043 NX}, 
%    country={The Netherlands}}

\author[2]{Jacob Jeffries} %[style=chinese]
\affiliation[2]{Department of Materials Science and Engineering, Clemson University, Clemson, SC, 29362, USA}
\author[2]{Muhammed Kose}
%[%
%   role=Co-ordinator,
%   suffix=Jr,
%\ead{cvr3@sayahna.org}
%\ead[URL]{www.sayahna.org}

\credit{Data curation, Writing - Original draft preparation}

%\affiliation[2]{organization={Sayahna Foundation},
%    city={Jagathy},
    % citysep={}, % Uncomment if no comma needed between city and postcode
%    postcode={695014}, 
%    state={Trivandrum},
%    country={India}}

\author%
[1,2]
{Enrique Martinez}
\cormark[1]
%\fnmark[1,3]
\ead{enrique@clemson.edu}
\ead[URL]{muexly}

%\affiliation[3]{organization={STM Document Engineering Pvt Ltd.},
%    addressline={Mepukada}, 
%    city={Malayinkil},
    % citysep={}, % Uncomment if no comma needed between city and postcode
%    postcode={695571}, 
%    state={Trivandrum},
%    country={India}}

\cortext[cor1]{Corresponding author}
%\cortext[cor2]{Principal corresponding author}

%\fntext[fn1]{This is the first author footnote. but is common to third
%  author as well.}
%\fntext[fn2]{Another author footnote, this is a very long footnote and
%  it should be a really long footnote. But this footnote is not yet
%  sufficiently long enough to make two lines of footnote text.}

%\nonumnote{This note has no numbers. In this work we demonstrate $a_b$
%  the formation Y\_1 of a new type of polariton on the interface
%  between a cuprous oxide slab and a polystyrene micro-sphere placed
%  on the slab.
%  }

\begin{abstract}
One key challenge for efficiency and safety in fusion devices is the retention of tritium (T) in plasma-facing components. Tritium retention generates radioactive concerns and decreases the amount of fuel available to generate power. Hence, understanding the behavior of T in tungsten (W), as the main candidate as armor material, is critical to the deployment of fusion as a reliable energy source. In this work, we have studied the effect of a thermal gradient in the transport properties of hydrogen (as a T surrogate) in pure W. Strong thermal gradients develop in the divertor as a result of the intense energy fluxes arriving at the material. We have developed an analytical approach to compute the heat of transport ($Q^*$) that is parameterized from molecular dynamics (MD) simulations. $Q^*$ is a parameter needed in irreversible thermodynamics frameworks to understand mass transport in the presence of thermal gradients. We show that $Q^*$ can be written as a function of temperature, temperature gradient, a characteristic length and the ratio of the rates towards hot and cold regions. Furthermore, we describe how, to first order, the dependence of $Q^*$ on the thermal gradient vanishes, in agreement with MD results. On average, we find $Q^*=\SI{-5.41e-3}kT^2{eV}$ for H in pure W, with $k$ the Boltzmann constant and $T$ the temperature.
\end{abstract}

% \includegraphics{figs/grabs.pdf}
% Keywords

\maketitle

\section{Introduction}

In the quest for developing nuclear fusion as a clean and sustainable source of energy, a critical component is the optimization of plasma-facing components (PFCs). These components are integral to the operation of fusion reactors, and face the formidable challenge of withstanding the intense environment inside the reactor. This environment includes not only high temperatures but also significant fluxes of heat, charged particles, and neutrons, all of which are pivotal in determining the overall plasma confinement and the efficiency of the fusion process \cite{knaster2016materials}.

Tungsten (W) has emerged as the primary material of choice for PFCs in nuclear fusion devices due to its exceptional properties, such as low sputtering yield, high thermal conductivity, and high melting point, along with lower tritium retention compared to other materials \cite{hirai2016use,roth2011hydrogen}
. The design of W monoblocks for the divertor targets, considering these properties and the anticipated harsh environment, is a testament to the complexity of the challenges in fusion technology. These monoblocks, with their specific geometry and cooling mechanisms, must withstand extreme thermal gradients, with surface temperatures potentially reaching over $2,300$ K while maintaining much lower temperatures at the cooling pipe, leading to gradients of about $317$ K/mm \cite{pitts2019physics}
.

A key aspect of this environment is the interaction of PFCs with hydrogen isotopes (deuterium, tritium), particularly at the divertor, where the plasma-wall interactions are most intense. The fluxes of these particles, on the order of \(10^{24} \, \text{m}^{-2}\text{s}^{-1}\), pose significant challenges to the integrity of the material surfaces. These interactions influence both the boundary and core plasma characteristics, and understanding, managing, and controlling them is essential for the success of fusion energy as a viable power source. Issues such as material erosion, tritium entrapment in redeposited layers, and high heat flux leading to melting and thermal fatigue are some of the critical challenges that need to be addressed \cite{roth2008tritium,doerner2019implications,murdoch1999tritium,schmid2012hydrogen,ogorodnikova2008ion}.
These plasma-wall interactions have been extensively investigated due to their critical impact on material integrity and plasma performance, with previous experimental studies highlighting tungsten’s resistance yet susceptibility under extreme plasma exposure conditions \cite{frauenfelder1969solution,philipps2011tungsten,wright2012hydrogenic}. Complementary to these studies, tritium transport modeling frameworks like TMAP have been developed to interpret retention data and predict inventory evolution in plasma-facing components \cite{longhurst2005verification}.
 Additionally, helium-induced damage can severely degrade thermal properties of tungsten, which affects T retention and thermal transport under divertor-like conditions \cite{hu2017thermal,wang2021helium,wirtz2016helium}.

Prior studies have indicated the significance of thermomigration, or the Soret effect, in these environments, with species tending to migrate towards hotter regions, potentially enriching these areas with T \cite{longhurst1985soret,baskes1982tritium,sugisaki1982isotope,markelj2020deuterium}. This phenomenon links a thermal gradient with the flux of matter and is vital in understanding T retention and permeation~\cite{asaro2007soret,dasgupta2023soret,heinola2010diffusion,schmid2017recent}. 

The focus of this paper is to deepen the understanding of how these large thermal gradients, along with the intense plasma-material interactions, affect the transport of atomic species such as hydrogen atoms (serving as surrogates for tritium), and consequently influence the microstructural evolution of PFC materials. This study is crucial as tritium is not only a key element in the fusion process but also a safety and biological hazard, necessitating precise models to predict its behavior and concentration within the reactor.

\section{Methodology}

Our investigation was centered around an in-depth examination of the diffusion patterns exhibited by a solitary hydrogen (H) atom within bulk W. Our primary focus was to determine the influence of a thermal gradient on these patterns. To accomplish this, we employed nonequilibrium molecular dynamics (NEMD) simulations. The simulations were conducted within a simulation box with dimensions of \(2.531 \times 2.531 \times 56.949 \, \text{nm}^3\) and orientation $x=[100]$, $y=[010]$, and $z=[001]$. Throughout the simulations, we ensured that periodic boundary conditions were maintained in all three dimensions (x, y, and z). We relied on an Embedded Atom Method potential to describe the interactions between atoms \cite{daw1993embedded,Marinica2013EAM}. According to this model, H atoms exhibit a preference for occupying tetrahedral sites within the body-centered cubic (BCC) lattice structure of W. 

The LAMMPS computational code was employed to execute the simulations \cite{plimpton1995fast}. The system is equilibrated at the average temperature and zero pressure for 10 ps. A consistent temperature was maintained in the central region of the specimen along the z-axis, which was the heat source, as well as at the periphery of the sample, that behaved as the heat sink, regulated by a Langevin thermostat with a characteristic time of 0.1 ps. Hence, the temperature at the center of the domain was higher than the temperature at the edges, generating a heat flux from the center to the edges. During the generation of the thermal gradient and the subsequent data collection, the barostat was removed to carry the measurements on an NVT ensemble. Two sets of simulations were performed. First, the average temperature was maintained constant at $1,300$ K and a varying temperature gradient was applied, from $17.54$ to $35.08$ K/nm. Second, a constant temperature gradient of $28.07$ K/nm was applied, modifying the average temperature spanning from $1,300$ to $1,800$ K. The goal was to identify the role of each component in the value of the heat of transport.

According to irreversible thermodynamics, the particle current $j$ is given as
\begin{equation}
   % j=-D\frac{\partial c}{\partial z} - D\frac{cQ^*}{kT^2}\frac{\partial T}{ \partial z}
    j=-D\nabla c - D\frac{cQ^*}{kT^2}\nabla T
    \end{equation}
where $c(z,t)$ denotes the concentration, $T(z,t)$ is the temperature field, $D$ is the diffusivity, $k$ is the Boltzmann constant, and $Q^*$ is the heat of transport, that couples the thermal gradient with the concentration profile.

At equilibrium, $j=0$, we can solve for the concentration profile
\begin{equation}
   % \frac{\partial c}{\partial z} = -\frac{cQ^*}{kT^2} \frac{\partial T}{\partial z}.
    \nabla c = -\frac{cQ^*}{kT^2} \nabla T.
\end{equation}

and integrating between $z_0$ and $z$
\begin{equation}\label{eq:profileIT}
    %c(z) = c(z_0) \exp \left( -\int_{z_0}^{z} \frac{Q^*}{kT^2}\frac{\partial T}{\partial z}dz \right).
    c(z) = c(z_0) \exp \left( -\int_{z_0}^{z} \frac{Q^*}{kT^2}\nabla T dz \right).
\end{equation}

\begin{figure}
    \centering
    \includegraphics[scale=.5]{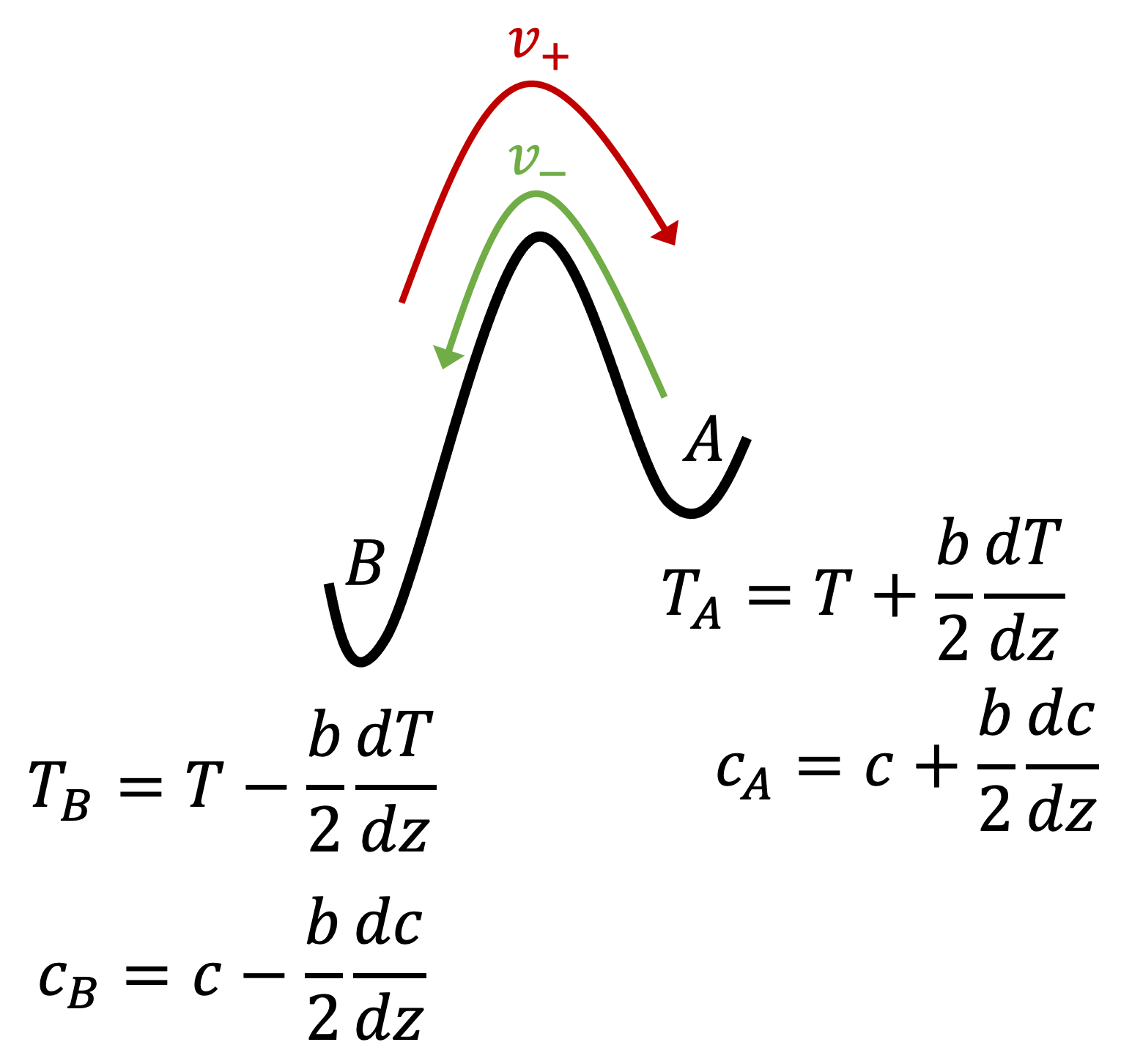}
    \caption{Hopping process for a particle in the presence of a thermal gradient. $v_+$ and $v_{-}$ represent the velocities towards the hot and cold regions, respectively.}
    \label{fig:jump}
\end{figure}

The particle current can also be written as
\begin{equation}\label{eq:current}
    j=c(z-\bar{b}/2,t)v^+ - c(z+\bar{b}/2,t)v^-,
\end{equation}
where $\bar{b}$ is a characteristic jump length and $v^+$ and $v^-$ are the velocities towards the hot and cold regions, respectively (see Fig.\ref{fig:jump}). We can therefore write
\begin{equation}
    v^+=\frac{\bar{b}\Delta N^+}{t_0}; \quad v^-=\frac{\bar{b}\Delta N^-}{t_0},
\end{equation}
with $\Delta N^+$ and $\Delta N^-$ the excess number of jumps to the hot and cold regions, respectively, and $t_0$ the time between observations. We can define
\begin{equation}
    \phi = \frac{v^+}{v^-}=\frac{\frac{\bar{b}\Delta N^+}{t_0}}{\frac{\bar{b}\Delta N^-}{t_0}}=\frac{\Delta N^+}{\Delta N^-},
\end{equation}
as the ratio between velocities to hot and cold regions, which can be estimated as the ratio of the number of jumps to hot and cold regions. Substituting $\phi$ in Eq. (\ref{eq:current}) and Taylor expanding the concentration we obtain at equilibrium
\begin{equation}
    \left ( c - \frac{\bar{b}}{2}\frac{dc}{dz} \right )\phi - \left ( c + \frac{\bar{b}}{2}\frac{dc}{dz} \right ) = 0,
\end{equation}

\begin{equation}
   - \left ( 1 -\phi \right )c - \left ( 1 + \phi \right )\frac{\bar{b}}{2}\frac{dc}{dz} = 0,
\end{equation}

\begin{equation}
    \frac{dc}{dz} = -\frac{2}{\bar{b}}\frac{\left ( 1 -\phi \right )}{\left ( 1 + \phi \right )}c.
\end{equation}
Integrating
\begin{equation}\label{eq:profileJumps}
    c(z) = c(z_0)\exp \left ( -\int _{z_0}^{z} \frac{2}{\bar{b}}\frac{\left ( 1 -\phi \right )}{\left ( 1 + \phi \right )}dz \right).
\end{equation}
Equating Eqs. (\ref{eq:profileIT}) and (\ref{eq:profileJumps})
\begin{align*}
    \exp \left ( -\int _{z_0}^{z} \frac{2}{\bar{b}}\frac{\left ( 1 -\phi \right )}{\left ( 1 + \phi \right )}dz \right) &= \exp \left( -\int_{z_0}^{z} \frac{Q^*}{kT^2}\nabla T dz \right), \\
    -\frac{2}{\bar{b}}\frac{\left ( 1 -\phi \right )}{\left ( 1 + \phi \right )} &= -\frac{Q^*}{kT^2}\nabla T.
\end{align*}
Therefore, the heat of transport can be written as
\begin{equation}
    Q^*(z) = \frac{2}{\bar{b}}\frac{\left ( 1 -\phi \right )}{\left ( 1 + \phi \right )}\frac{kT(z)^2}{\nabla T}.
\end{equation}
Noting that the effect of the thermal gradient can be treated as a perturbation, we can further Taylor expand $\frac{1-\phi}{1+\phi}$ around $\phi = 1$, which to second order gives
\begin{equation}
    Q^*(z) = \frac{2}{\bar{b}}\frac{kT(z)^2}{\nabla T}\left ( -\frac{1}{2}(\phi - 1)+\frac{1}{4}(\phi - 1)^2 \right ),
\end{equation}
and to first order
\begin{equation}\label{eq:alpha}
    Q^*(z) = \frac{1}{\bar{b}}\frac{kT(z)^2}{\nabla T}\left ( 1 -\phi  \right )=\alpha kT^2(z).
\end{equation}
This expression highlights the dependence of $Q^*$ on the local temperature $T(z)$, the temperature gradient $\nabla T(z)$, the velocity ratio $\phi$ and the characteristic length $\bar{b}$, and where we have defined $\alpha=\frac{2}{\bar{b}\nabla T} \frac{\left ( 1-\phi \right )}{\left ( 1+\phi \right )}\approx \frac{(1-\phi)}{\bar{b}\nabla T}$.

To test the expressions above, we investigated the dynamics of transport in the presence of thermal gradients of one hydrogen atom starting at a random tetrahedral location and executed five independent simulation runs of equal duration, each lasting 150 ns. The positions of the hydrogen atoms were meticulously monitored at intervals of 0.5 ps. Subsequently, the concentration of these atoms was computed by dividing the sample along the z-axis (direction of the thermal gradient) into segments that were 1 nm wide. Through this segmentation, we were able to ascertain the proportion of time that each atom occupied within these segments in relation to their respective volumes, obtaining the concentration profiles.

This investigation was intentionally devised to concentrate on situations where there are minimal quantities of defects (dilute assumption), thus avoiding the clustering of defects by restricting the simulations to individual hydrogen atoms within the MD cells. Interactions between hydrogen atoms are mainly repulsive and therefore hydrogen clustering is negligible, implying that the heat of transport of a single hydrogen atom is the most important value to study hydrogen retention. However, synergies between hydrogen and helium are also of importance, although they will be studied in a future work, particularly in light of previous studies demonstrating mobile helium cluster dynamics and surface interactions in tungsten \cite{hu2014helium, hu2015mdhelium}. Also density functional theory studies reveal that hydrogen interacts strongly with vacancies in tungsten, which can significantly alter its diffusion pathways \cite{fernandez2015dftdiffusion}. 
%This methodology is of utmost importance, particularly when considering the possibility of hydrogen clustering in tungsten, which is similar to the observed behavior with helium. The main objective of this research was to establish a multiscale modeling framework that is capable of computing the heats of transport and to explore the influence of the Soret effect on the retention of hydrogen within PFC tungsten.

%This comprehensive analysis elucidates the dynamics of thermally induced migration of hydrogen, especially pertinent to plasma-facing materials in fusion reactors. It highlights the profound impact of thermal gradients on the performance of materials in the challenging environments of fusion energy systems.
%\bibliographystyle{unsrt}
\section{Results}

\subsection{Effect of $\nabla T$}

\begin{figure}
    \centering
    \includegraphics[scale=.5]{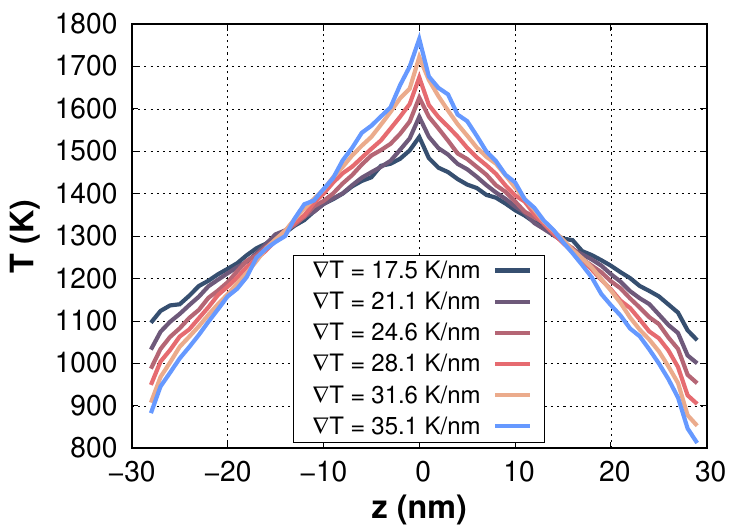}
    \caption{Steady-state temperature profiles for different applied thermal gradients as obtained from MD.}
    \label{fig:gradT}
\end{figure}

Figure \ref{fig:gradT} presents the average temperature profile among the five independent runs for each of the applied conditions. In all cases, we see that the profile is linear except in a region close to the applied thermostats. It is worth noting that these temperature profiles are obtained from classical MD simulations without any electronic contribution, although W is a metal with high thermal conductivity. However, it is experimentally observed that the thermal conductivity does not depend strongly on the temperature in the thermal range studied in this work and can be considered constant, which would not significantly modify the temperature profiles obtained here. Nonetheless, there could be a component due to electron-phonon interaction that we admittedly neglect in this work. Motivated by the observed linearity in the temperature profile, in what follows we consider the temperature gradient equal to the applied gradient.

\begin{figure}
    \centering
    \includegraphics[scale=.5]{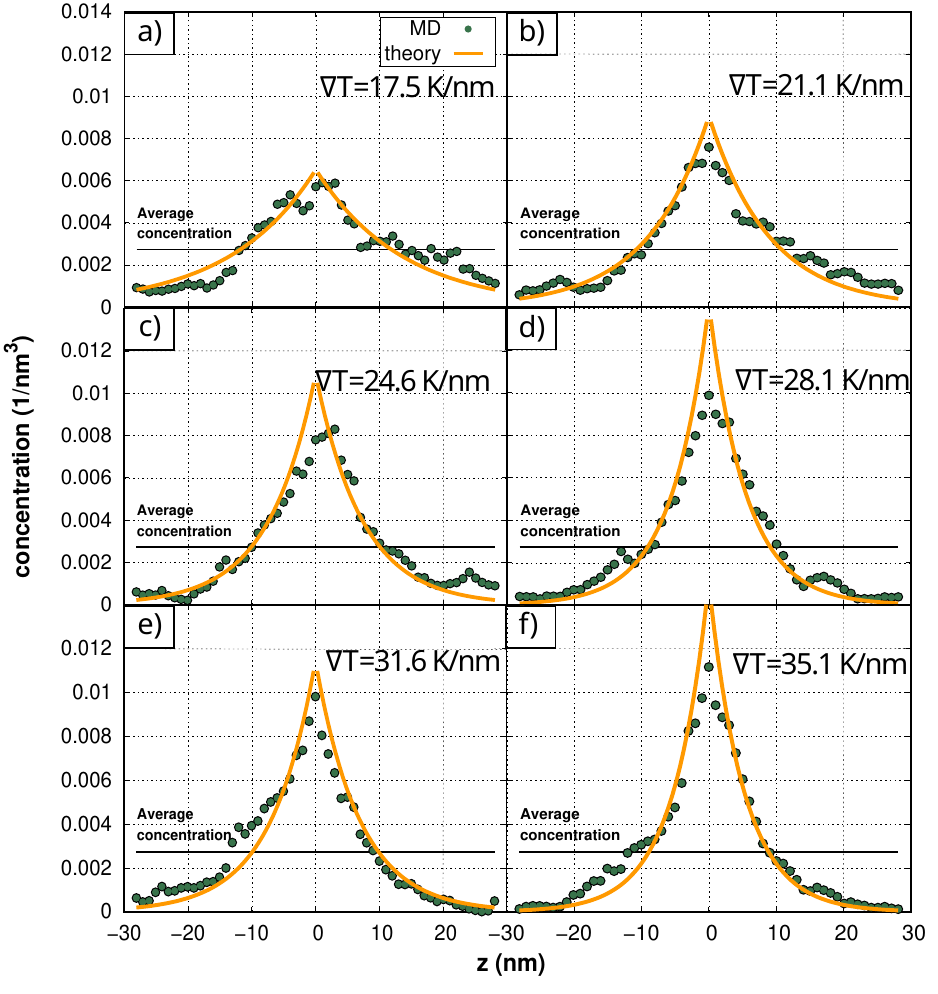}
    \caption{Hydrogen concentration profiles depending on the applied thermal gradient for the same average temperature ($1,300$ K) as given by non-equilibrium MD and Eq.~\ref{eq:profileJumps} with the parameters shown in Table \ref{tab:simulation-results}.}
    \label{fig:gradTH}
\end{figure}

Figure \ref{fig:gradTH} shows the H concentration profiles at steady state in the presence of different temperature gradients with the same average temperature (1300 K) as given by NEMD and Eq.~\ref{eq:profileJumps} with the parameters obtained from the atomistic trajectory, presented in Table ~\ref{tab:simulation-results}. We observe that in all cases the propensity for H to stay around the hotter region of the sample is higher than at the cold region, as it was reported in previous publications \cite{martinez2021thermal,dasgupta2023soret}. To improve the statistical analysis, we have further subdivided each of the five independence trajectories in ten equally long subtrajectories to compute the means and standard deviations. 

The jump length of H tetrahedral in the $z$ direction would be $a_0(T)/4$, where $a_0(T)$ is the lattice parameter of W. At $0$ K, the lattice parameter of W as predicted by the interatomic potential is $a_0(0)=0.314$ nm. Hence, the jumping distance at $0$ K without considering any strain induced by the H atom would be $\delta _0 = 0.0785$ nm, similar to the values we measured in the MD runs as the characteristic jump length $\bar{b}$. We have computed the thermal expansion coefficient for the interatomic potential used in this work in the temperature range considered, giving $\alpha _T=1.421\times 10^{-5}$ 1/K. The hopping length for the H was then obtained as $\delta = \delta _0 \left ( 1 +\alpha _T \Delta T \right )$. From the NEMD, we measure the total distance traveled by the H after 0.5 ps. If the distance is larger than $0.75\delta$ it was assumed that the H jumped. Furthermore, if the measured distance is larger than $2\delta$ the jump is considered correlated and counted just as one. Taking the activation barrier as $\Delta H = \SI{0.38}{eV}$ and the prefactor $\nu = \SI{4.43e13}{s^{-1}}$ \cite{dasgupta2023soret}, we estimate that the probability for the H to hop twice in the same direction under thermal conditions (without a thermal gradient) is less than 12.5\% at the maximum local temperature used in this study, 2200 K. At 1000 K the probability decreases to around 5\%. Once we have the total number of jumps to the hot and cold regions, we estimate $\phi$ as their ratio.

%The characteristic jump distance $\bar{b}$ was 

%computed following the same scheme, computing the average of the jumps after 0.5 ps and then averaging over the same subtrajectories. In both cases, a jump is considered if the distance is larger than 0.005 nm. Note, however, that in this work we are not tracking every thermally activated jump of the H atom, but we are considering its displacements in the whole continuum space.

To compute $c_0$, the concentration at the origin, we impose the condition that the average concentration is the same as in MD, i.e., one H atom over the volume of the system. Since the setup is symmetric the following expression needs to hold

\begin{equation}
    \begin{split}
    \langle c \rangle & = \frac{1}{L_z}\int _{0}^{L_z} c(z)dz \\
    & =\frac{1}{L_z}\int _{0}^{L_z} c(z_0) \exp \left ( \int _{0} ^{z} -\frac{2(1-\phi)}{\bar{b}(1+\phi)}dz' \right ) dz.
    \end{split}
\end{equation}

\noindent Considering the term $\frac{2(1-\phi)}{\bar{b}(1+\phi)}$ as independent of $z$, we obtain the solution
\begin{equation}
    \langle c \rangle = \frac{c(z_0)}{\zeta}\left [ \exp \left ( -\zeta \right ) -1 \right],
\end{equation}
where 
\begin{equation}
    \zeta =\frac{2L_z(1-\phi)}{\bar{b}(1+\phi)},
\end{equation}
to obtain $c(z_0)$, which values are reported in Tables \ref{tab:simulation-results} and \ref{tab:simulation-results-meanT}.

%\begin{table}[H]  %Geomtric b
%    \centering
%    \begin{tabular}{|c|c|c|c|c|}
%    \hline
%    $\nabla T$ (K/nm) & $\phi$ & $\bar{b}$ (nm) & $c(z_0)$ (nm$^{-3}$) & $\alpha$ (K$^{-1}$) \\
%    \hline
%    17.5 & 1.00604 & 0.07995 & 0.00664 & -0.004316 \\
%    \hline
%    21.1 & 1.00878  & 0.07995  & 0.00891 & -0.005204 \\
%    \hline
%    24.6 & 1.00755 & 0.07995 & 0.00786 & -0.003838 \\
%    \hline 
%    28.1 & 1.01294  &  0.07995 &  0.01265 & -0.005759 \\
%    \hline
%    31.6 & 1.01074 & 0.07995 & 0.01064 & -0.004251 \\
%    \hline
%    35.1 & 1.01416 & 0.07995 & 0.01378 & -0.005046 \\
%    \hline
%    \end{tabular}
%    \caption{Mean  values for $\phi$, $\bar{b}$, $c_0$, and $\alpha$ as a function of the applied $\nabla T$ from MD for the same average temperature of $1,300$ K.}
%    \label{tab:simulation-results}
%\end{table}

\begin{table}[H]
    \centering
    \begin{tabular}{|c|c|c|c|c|}
    \hline
    $\nabla T$ (K/nm) & $\phi$ & $\bar{b}$ (nm) & $c(z_0)$ (nm$^{-3}$) & $\alpha$ (K$^{-1}$) \\
    \hline
    17.54 & 1.00589 & 0.07995 & 0.00653 & -0.00420 \\
    \hline
    21.05 & 1.00895  & 0.07995  & 0.00906 & -0.00528 \\
    \hline
    24.56 & 1.01103 & 0.07995 & 0.01090 & -0.00558 \\
    \hline 
    28.07 & 1.01455  &  0.07995 &  0.01415 & -0.00643 \\
    \hline
    31.57 & 1.01163 & 0.07995 & 0.01145 & -0.00458 \\
    \hline
    35.08 & 1.01548 & 0.07995 & 0.01503 & -0.00547 \\
    \hline
    \end{tabular}
    \caption{Mean  values for $\phi$, $\bar{b}$, $c_0$, and $\alpha$ as a function of the applied $\nabla T$ from MD for the same average temperature of $1,300$ K.}
    \label{tab:simulation-results}
\end{table}

The theoretical solution, solid orange lines in Fig.~\ref{fig:gradTH}, follows closely the observations from MD, with, in general, steeper H concentration profiles for larger temperature gradients, which serves as validation for the theoretical approach.

\subsection{Effect of mean temperature}
\begin{figure}
    \centering
    \includegraphics[scale=.5]{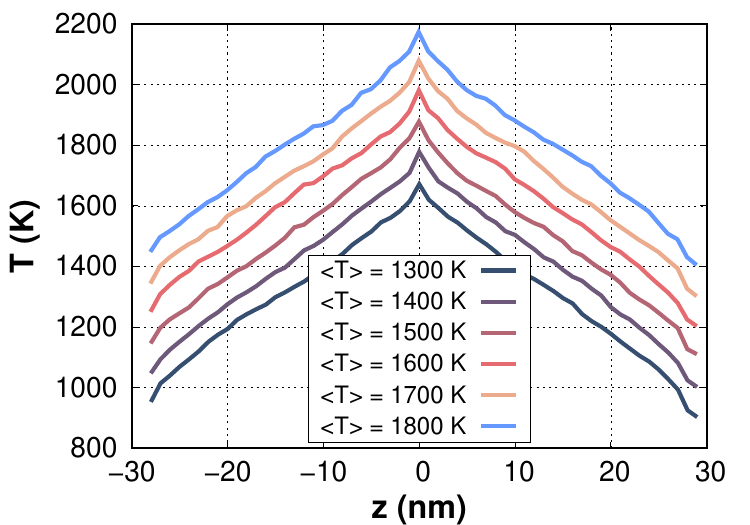}
    \caption{Steady-state temperature profiles for different applied thermal gradients as obtained from MD.}
    \label{fig:gradTmean}
\end{figure}

Figure \ref{fig:gradTmean} shows the temperature profile for the different cases studied, with the same thermal gradient applied and different average temperatures. We have analyzed average temperatures from 1300 K to 1800 K, holding the gradient constant at 28.07 K/nm. As for the cases described above with similar mean temperature and varying thermal gradients, the temperature profiles are mainly linear but in a region close to the thermostated domains.

\begin{figure}
    \centering
    \includegraphics[scale=.40]{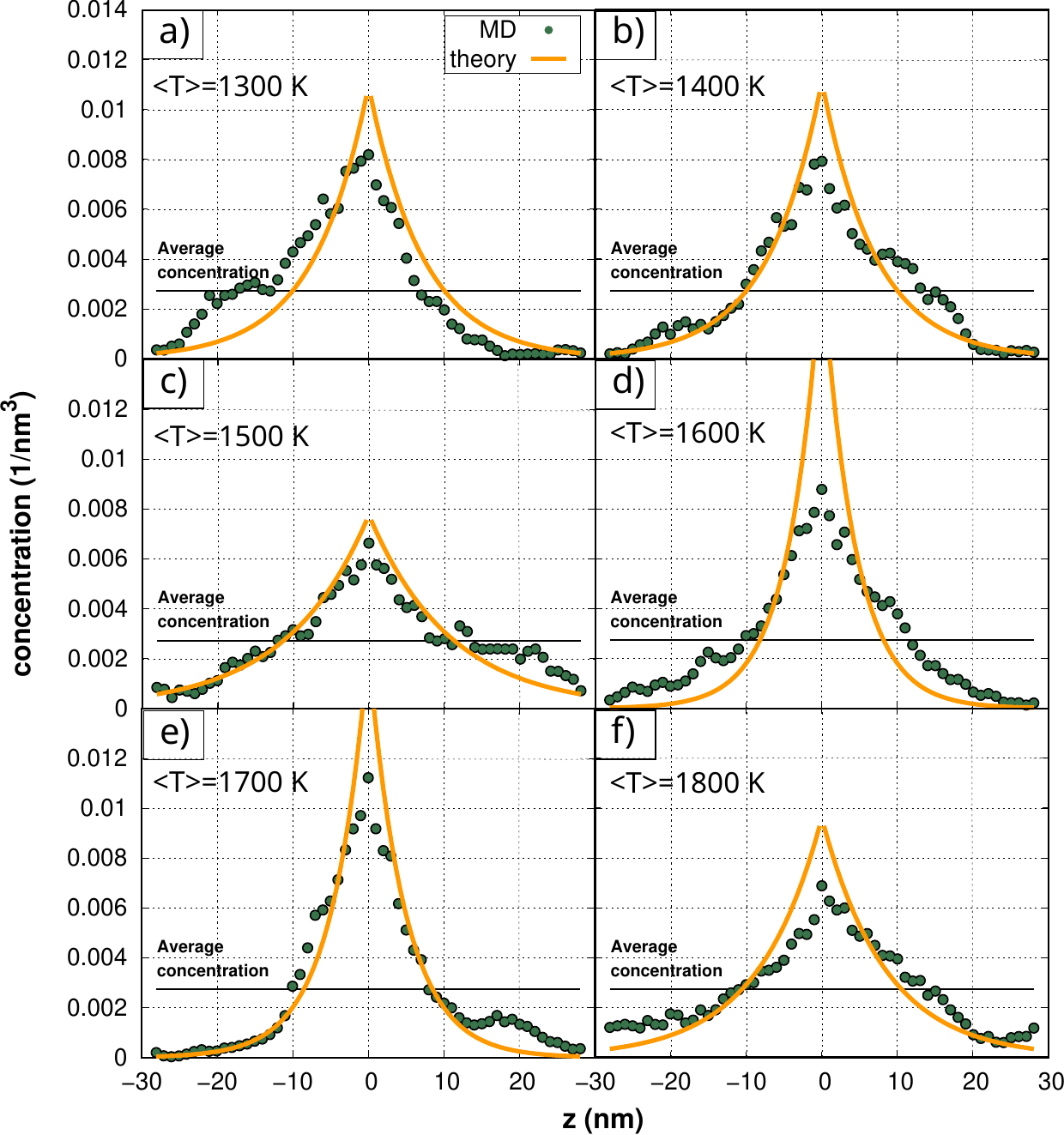}
    \caption{Hydrogen concentration profiles depending on the average temperature for the same applied thermal gradient (28.07 K/nm) as given by non-equilibrium MD and Eq.~\ref{eq:profileJumps} with the parameters shown in Table \ref{tab:simulation-results-meanT}.}
    \label{fig:gradTmeanH}
\end{figure}

We have also compared the H concentration profiles at steady-state for these cases with similar thermal gradients and varying mean temperatures. The results are shown in Fig.~\ref{fig:gradTmeanH}, where the theoretical approach described above (solid orange lines) is compared with the results from NEMD simulations (green points). The profiles obtained from both methodologies agree remarkably well. $\phi$ is again computed similarly, as the ratio of the number of jumps towards the hot region over the number of jumps in the direction of the cold region. These measurements were done with the same subtrajectory strategy described above.

Table \ref{tab:simulation-results-meanT} summarizes the values for the parameters measured during the NEMD simulations.
%\begin{table}[H]  %b from geometry
%    \centering
%    \begin{tabular}{|c|c|c|c|c|}
%    \hline
%    $\langle T \rangle$ (K) & $\phi$ & $\bar{b}$ (nm) & $c(z_0)$ (nm$^{-3}$) & $\alpha$ (K$^{-1}$) \\
%    \hline
%    1,300 & 1.00853 & 0.07995 & 0.008695 & -0.0050564 \\
%    \hline
%    1,400 & 1.00959  & 0.080061  & 0.009606 & -0.0056769 \\
%    \hline
%    1,500 & 1.00745 & 0.080173  & 0.007765 & -0.0044039 \\
%    \hline
%    1,600 & 1.0112 & 0.08028  & 0.01101 & -0.0066115 \\
%    \hline
%    1,700 & 1.01366 & 0.08039 & 0.01325 & -0.008052 \\
%    \hline
%    1,800 & 1.00712 & 0.080507 & 0.007469 & -0.004191 \\
%    \hline
%    \end{tabular}
%    \caption{Mean  values for $\phi$, $\bar{b}$, $c_0$, and $\alpha$ as a function of the applied average temperature $\langle T \rangle$ from MD for the same $\nabla T = 21.1$ K/nm.}
%    \label{tab:simulation-results-meanT}
%\end{table}

\begin{table}[H]
    \centering
    \begin{tabular}{|c|c|c|c|c|}
    \hline
    $\langle T \rangle$ (K) & $\phi$ & $\bar{b}$ (nm) & $c(z_0)$ (nm$^{-3}$) & $\alpha$ (K$^{-1}$) \\
    \hline
    1,300 & 1.0111 & 0.07995 & 0.01097 & -0.00491 \\
    \hline
    1,400 & 1.0113  & 0.08006  & 0.01113 & -0.004995 \\
    \hline
    1,500 & 1.00745 & 0.08017  & 0.00777 & -0.00329 \\
    \hline
    1,600 & 1.01856 & 0.08028  & 0.01789 & -0.00815 \\
    \hline
    1,700 & 1.01699 & 0.08039 & 0.01639 & -0.00746 \\
    \hline
    1,800 & 1.00962 & 0.08051 & 0.00960 & -0.00423 \\
    \hline
    \end{tabular}
    \caption{Mean  values for $\phi$, $\bar{b}$, $c_0$, and $\alpha$ as a function of the applied average temperature $\langle T \rangle$ from MD for the same $\nabla T = 28.07$ K/nm.}
    \label{tab:simulation-results-meanT}
\end{table}

\section{Discussion}\label{sec:discussion}
We are discussing here the dependence of the parameters involved in the equations to compute $Q^*$, mainly $\phi$, and $\alpha$, with the temperature and the temperature gradient. The goal is to understand the conditions under which these parameters must be computed to estimate $Q^*$.

\subsection{Dependence of $\phi$ on $\langle T \rangle$ and $\nabla T$}
\begin{figure}
    \centering
    \includegraphics[width=\linewidth]{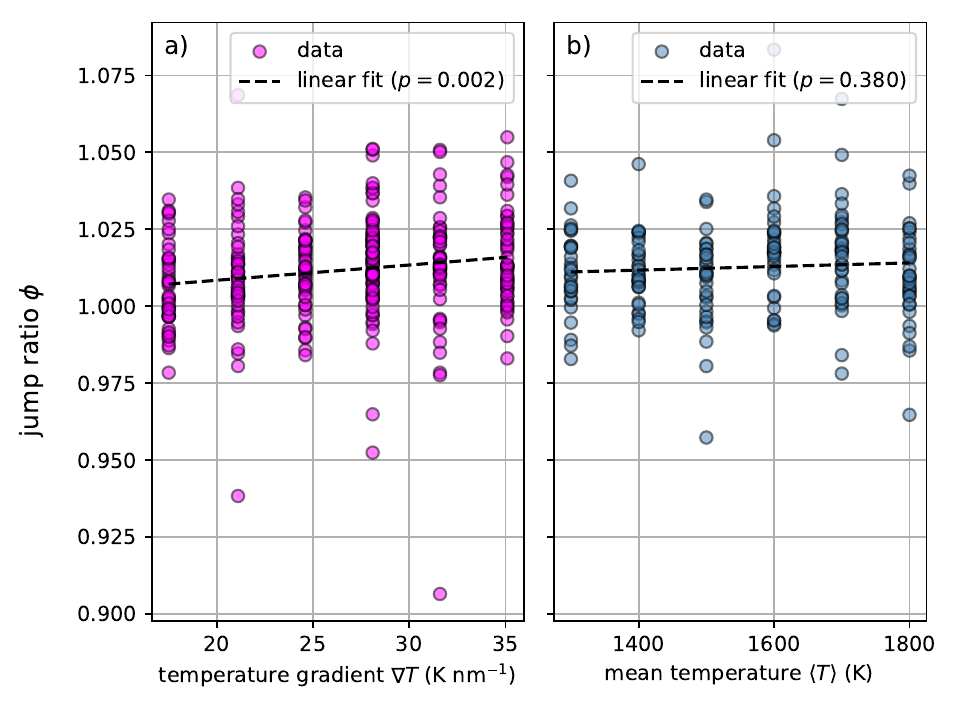}
    \caption{Computed NEMD values (points) and linear fits (dashed lines) for $\phi$ in terms of $\nabla T$ (a) and $\langle T\rangle$ (b), and corresponding $p$-values for the null hypothesis that, respectively, $\beta = 0$ and $\delta = 0$.}
    \label{fig:hypothesis-testing-phi}
\end{figure}

From Tables \ref{tab:simulation-results} and \ref{tab:simulation-results-meanT} the dependence of $\phi$ on either $\nabla T$ or $\langle T \rangle$ is subtle. In the case of the dependence with $\nabla T$, an increase in values for stronger gradients seems to be present. The dependence with $\langle T \rangle$ seems to be weaker, without a clear trend as we vary the average temperature. Figure \ref{fig:hypothesis-testing-phi} shows the dependence of the values computed for $\phi$ as a function of the thermal gradient $\nabla T$ (Fig.~\ref{fig:hypothesis-testing-phi}(a)) and the average temperature $\langle T \rangle$ (Fig.~\ref{fig:hypothesis-testing-phi}(b)). We used both a linear function and a constant value to fit the MD results, with linear functional forms $\phi(\nabla T)=1+\beta \nabla T$ and $\phi(\langle T \rangle )=1+\delta \langle T \rangle$. For $\phi(\nabla T)$, the linear fit gives a value $\beta =0.0003834$ nm/K with a root mean square residual (RMSR) of $0.001997$, while using a constant value $\phi = 1.010035$ the RMSR is $0.003123$. Hence, there is a significant difference in the error from both approaches, with the linear fit being more accurate than the constant value. On the other hand, for the case of the dependence on $\langle T \rangle$, we obtain a value of $\delta =6.3738 \times 10^{-6}$ 1/K, with an RMSR $0.002979$ and for the constant $\phi$ we gather a value $\phi = 1.009592$ with an RMSR equal to $0.003009$, very similar to the error obtained from the linear fitting. 

We have additionally performed $t$-tests for linear regression \cite{2020SciPy-NMeth} on the NEMD data with the null hypotheses that $\beta = 0$ or $\delta = 0$. Figure \ref{fig:hypothesis-testing-phi} shows the results of these hypothesis tests, yielding respective $p$-values of $0.002$ and $0.380$. For a confidence level of $0.01$, we can therefore confidently reject the hypothesis that $\beta = 0$ since $p < 0.01$, but we cannot confidently reject the hypothesis that $\delta = 0$ since $p > 0.01$.

Hence, we can conclude that $\phi$ weakly depends on the mean temperature, but more strongly depends on the temperature gradient. We can write then, to first order, $\phi (\nabla T )=1+ \beta \nabla T$. It seems logical that this is the case since the average temperature should not modify the propensity for H to jump on either direction, and it is the temperature gradient the term that adds a bias in the process, the driving force for H to jump towards the hot region more often than to the cold region. If we substitute the linear relation between $\phi$ and $\nabla T$ in Eq. \ref{eq:alpha}, we obtain 

\begin{align}\label{eq:alpha_new}
    Q^*(z) &= \frac{1}{\bar{b}}\frac{kT(z)^2}{\nabla T}\left ( 1 - \left ( 1 + \beta \nabla T \right )  \right ) \nonumber \\
   &= \frac{-\beta}{\bar{b}}kT(z)^2 = \alpha kT^2(z).
\end{align}
where $\alpha = -\beta/\bar{b}$ denotes the ratio between the slope of $\phi$ with respect to $\nabla T$ and the characteristic jump length $\bar{b}$. The sign of $\beta$ then determines the sign of $\alpha$ and thus the sign of $Q^*$. If $\beta > 0$ then $\alpha < 0$ and $Q^* < 0$ and the jumping specie tends to flow towards the hot region, while for $\beta < 0$, $\alpha > 0$ and $Q^* > 0$ and the flux tends to go to the cold regions. 
%\begin{figure}
%    \centering
%    \includegraphics[scale=.40]{figs/phiGratT-phiTmean.pdf}
%    \caption{Computed values of $\phi$ as a function of a) the temperature gradient $\nabla T$, and b) the average temperature $\langle T \rangle$. Green dots with errorbars denoting standard error represent the NEMD results. The solid green line illustrate a linear fit to the NEMD data, while the dashed blue line shows the best fit to a constant value.}
%    \label{fig:phi}
%\end{figure}

The origin of $\beta$ relates to phonon scattering mechanisms that can be analyzed with a Boltzmann transport approach. Relying on the dynamic theory of diffusion, Schottky derived an expression for the hopping rate of an atom into a vacant site as a function of a characteristic relaxation time $\tau$ and the thermal gradient for the specific process \cite{schottky1963thermal}. Current work focuses on deriving an analytical expression for $\beta$ as a function of the relaxation time $\tau$ and the phonon modes involved in the process.

\subsection{Dependence of $\alpha$ on $\langle T\rangle$ and $\nabla T$}

\begin{figure}
    \centering
    \includegraphics[width=\linewidth]{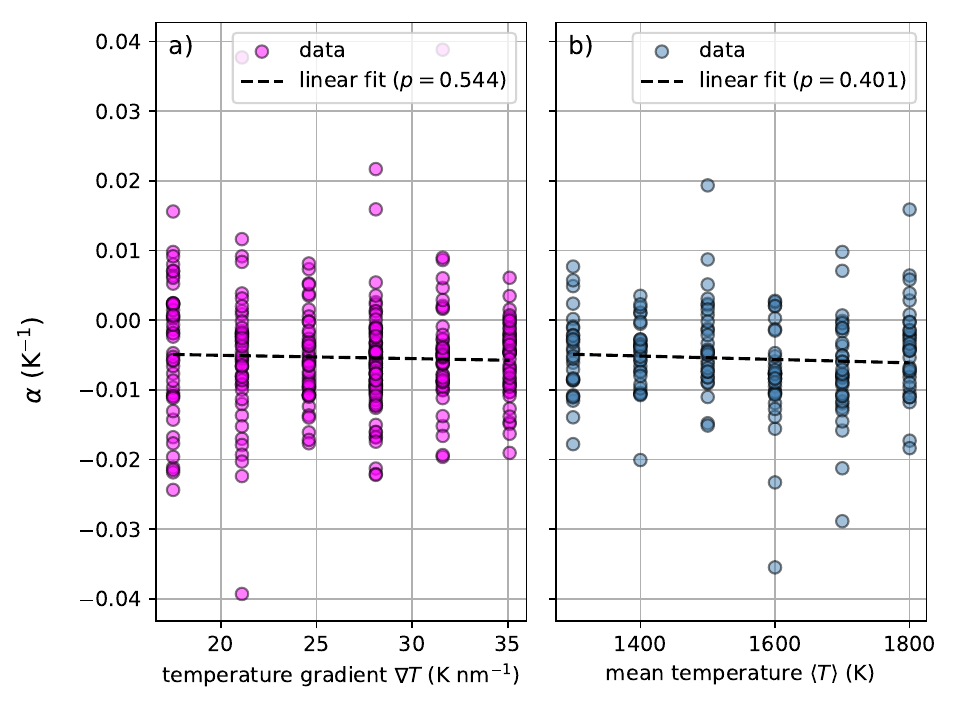}
    \caption{Computed NEMD values (points) and linear fits (dashed lines) for $\alpha$ in terms of $\nabla T$ (a) and $\langle T\rangle$ (b) and corresponding $p$-values for the hypotheses that either slope is $0$.}
    \label{fig:alpha-hypotheses}
\end{figure}

We have also analyzed the dependence of the ratio $\alpha$ on $\langle T \rangle$ and $\nabla T$. $\alpha$ seems to be largely independent of both $\nabla T$ and $\langle T\rangle$. We have used the same fitting procedure as described above for the values of $\phi$, with linear and constant functions to relate $\alpha$ to $\nabla T$ and $\langle T \rangle$. For the dependence with the thermal gradient we obtain an RMSR of $0.00084$ for the linear fit and $0.00080$ for the constant value. Hence, the constant value seems slightly more appropriate. Similarly, for the dependence of $\alpha$ with $\langle T\rangle$ we obtain an RMSR of $0.0020$ for the linear fit and $0.0019$ for the constant fit. Again, the constant value would deem more accurate.

Furthermore, we have performed $t$-tests for linear regression similar to the analysis for $\delta$ and $\beta$. Figure \ref{fig:alpha-hypotheses} shows the results of these hypothesis tests. The important result of this analysis is that the $p$-values for both $\alpha$'s dependence on $\nabla T$ and on $\langle T\rangle$ are quite large: respectively $0.544$ and $0.401$. For any reasonable confidence level, we cannot reject the hypotheses that either of these slopes are $0$. Therefore, we can conclude that $\alpha$'s dependence on both $\nabla T$ and $\langle T\rangle$ is weak at best, and hence $\alpha$ can be considered constant.

\begin{figure}
    \centering
    \includegraphics[width=\linewidth]{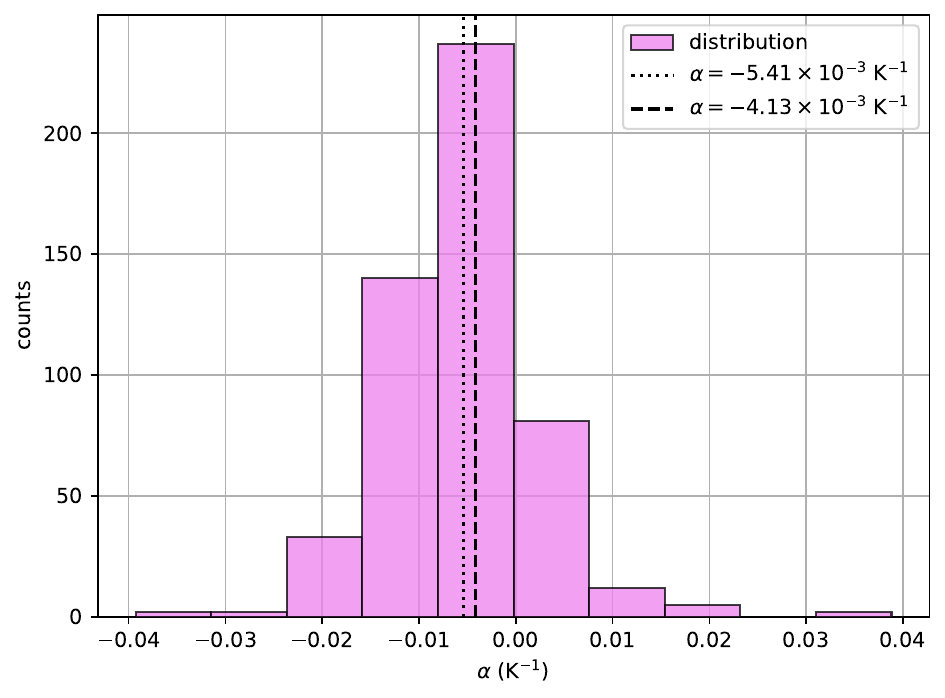}
    \caption{Distribution of $\alpha$ values as obtained by NEMD, showing the mode (dashed line) and the mean (dotted line).}
    \label{fig:alpha-distro}
\end{figure}

Finally, Fig. \ref{fig:alpha-distro} presents the distribution of $\alpha$ values obtained from NEMD. The distribution seems unimodal with mean $\SI{-5.41e-3}{K^{-1}}$ and mode $\SI{-4.13e-3}{K^{-1}}$.

%This dependence relates to the thermal expansion coefficient. The computed thermal expansion coefficient for the interatomic potential used in this work in the temperature range considered gives $\alpha _T=1.421\times 10^{-5}$ 1/K. A dashed gray line shows the trend that results using the computed thermal expansion coefficient, giving an RMSR of $0.00512$. The tendency is correct but the slope seems to be low compared to the NEMD results, probably due to size effects and boundary conditions constrains. 

%Since the dependence of $\bar{b}$ on $\langle T \rangle$ is significantly stronger than on $\nabla T$, to first order we could neglect its dependence on $\nabla T$ and estimate $\bar{b}$ as a linear function on $\langle T \rangle$. Furthermore, meaningful trends can still be obtained taking the of However, meaningful physical trends could be obtained to estimate the dependence of $\bar{b}$ on 

% Although the average temperature is the same, $\bar{b}$ seems to slightly increase with $\nabla T$. This is probably due to the thermal strains developed in the system, which are considered here as a component in $Q^*$. 

%b depends on T because the number of jumps increases, i.e., the diffusivity increases.

\section{Conclusions}\label{sec:conclusions}
Fusion imposes extreme environments to plasma facing materials (PFM), withstanding high thermal gradients and large influx of particles such as tritium and helium. In this paper we have studied the effect of thermal gradients on the migration of hydrogen, as a surrogate for tritium. We have developed an analytical expression for the heat of transport $Q^*$ (responsible for the coupling between thermal gradients and diffusion) that depends on temperature, the temperature gradient, a characteristic hopping length, and the ratio between the jump rates to the hot region over the cold region, $\phi$. We have tested such expression comparing to non-equilibrium molecular dynamics simulations of a hydrogen atom migrating in tungsten in the presence of thermal gradients. We have analyzed the dependence of $Q^*$ on the mean temperature and the temperature gradient and observed that $\phi$ depends linearly on the thermal gradient but is independent of the mean temperature, which, to first order, cancels the dependence of the heat of transport on the thermal gradient. Hence, we can write $Q^*=\alpha k T^2$, with $\alpha$ a constant. The analysis of the distribution of $\alpha$ values gives a mean value $\langle \alpha \rangle = \SI{-5.41e-3}{K^{-1}}$ and a mode $\alpha _{mode}=\SI{-4.13e-3}{K^{-1}}$. These developments will help consider the effects of thermal gradients on tritium retention and microstructure evolution in PFM. 

\section{Acknowledgments}\label{sec:acknowledgments}
Discussions with S. Blondel, D. Dasgupta, D. Maroudas and B. Wirth are gratefully acknowledged. The authors acknowledge support by the U.S.
Department of Energy, Office of Science, EPSCOR program
under Award Number DE-SC-0023385.

Additionally, this material is based on work supported by the National Science Foundation under Grant Nos. MRI\# 2024205, MRI\# 1725573, and CRI\# 2010270 for allotment of compute time on the Clemson University Palmetto Cluster.

\bibliographystyle{model1-num-names}
%bibliographystyle{cas-model2-names}

\bibliography{references}

\begin{thebibliography}{33}
\expandafter\ifx\csname natexlab\endcsname\relax\def\natexlab#1{#1}\fi
\providecommand{\url}[1]{\texttt{#1}}
\providecommand{\href}[2]{#2}
\providecommand{\path}[1]{#1}
\providecommand{\DOIprefix}{doi:}
\providecommand{\ArXivprefix}{arXiv:}
\providecommand{\URLprefix}{URL: }
\providecommand{\Pubmedprefix}{pmid:}
\providecommand{\doi}[1]{\href{http://dx.doi.org/#1}{\path{#1}}}
\providecommand{\Pubmed}[1]{\href{pmid:#1}{\path{#1}}}
\providecommand{\bibinfo}[2]{#2}
\ifx\xfnm\relax \def\xfnm[#1]{\unskip,\space#1}\fi
%Type = Article
\bibitem[{Knaster et~al.(2016)Knaster, Moeslang, and Muroga}]{knaster2016materials}
\bibinfo{author}{J.~Knaster}, \bibinfo{author}{A.~Moeslang}, \bibinfo{author}{T.~Muroga},
\newblock \bibinfo{title}{Materials research for fusion},
\newblock \bibinfo{journal}{Nature Physics} \bibinfo{volume}{12} (\bibinfo{year}{2016}) \bibinfo{pages}{424--434}.
%Type = Article
\bibitem[{Hirai et~al.(2016)Hirai, Panayotis, Barabash, Amzallag, Escourbiac, Durocher, Merola, Linke, Loewenhoff, Pintsuk, Wirtz, and Uytdenhouwen}]{hirai2016use}
\bibinfo{author}{T.~Hirai}, \bibinfo{author}{S.~Panayotis}, \bibinfo{author}{V.~Barabash}, \bibinfo{author}{C.~Amzallag}, \bibinfo{author}{F.~Escourbiac}, \bibinfo{author}{A.~Durocher}, \bibinfo{author}{M.~Merola}, \bibinfo{author}{J.~Linke}, \bibinfo{author}{T.~Loewenhoff}, \bibinfo{author}{G.~Pintsuk}, \bibinfo{author}{M.~Wirtz}, \bibinfo{author}{I.~Uytdenhouwen},
\newblock \bibinfo{title}{Use of tungsten material for the iter divertor},
\newblock \bibinfo{journal}{Nuclear Materials and Energy} \bibinfo{volume}{9} (\bibinfo{year}{2016}) \bibinfo{pages}{616--622}.
%Type = Article
\bibitem[{Roth and Schmid(2011)}]{roth2011hydrogen}
\bibinfo{author}{J.~Roth}, \bibinfo{author}{K.~Schmid},
\newblock \bibinfo{title}{Hydrogen in tungsten as plasma-facing material},
\newblock \bibinfo{journal}{Physica Scripta} \bibinfo{volume}{2011} (\bibinfo{year}{2011}) \bibinfo{pages}{014031}.
%Type = Article
\bibitem[{Pitts et~al.(2019)Pitts, Carpentier-Chouchana, Escourbiac, Hirai, Komm, Loarte, Merola, Mitrishkin, Panayotis, Riccardo, and You}]{pitts2019physics}
\bibinfo{author}{R.~Pitts}, \bibinfo{author}{S.~Carpentier-Chouchana}, \bibinfo{author}{F.~Escourbiac}, \bibinfo{author}{T.~Hirai}, \bibinfo{author}{M.~Komm}, \bibinfo{author}{A.~Loarte}, \bibinfo{author}{M.~Merola}, \bibinfo{author}{Y.~Mitrishkin}, \bibinfo{author}{S.~Panayotis}, \bibinfo{author}{V.~Riccardo}, \bibinfo{author}{J.-H. You},
\newblock \bibinfo{title}{Physics basis for the first iter tungsten divertor},
\newblock \bibinfo{journal}{Nuclear Materials and Energy} \bibinfo{volume}{20} (\bibinfo{year}{2019}) \bibinfo{pages}{100696}.
%Type = Article
\bibitem[{Roth et~al.(2008)Roth, Tsitrone, Loarte, Loarer, Causey, Sakamoto, Wampler, and Wilson}]{roth2008tritium}
\bibinfo{author}{J.~Roth}, \bibinfo{author}{E.~Tsitrone}, \bibinfo{author}{A.~Loarte}, \bibinfo{author}{T.~Loarer}, \bibinfo{author}{R.~Causey}, \bibinfo{author}{K.~Sakamoto}, \bibinfo{author}{W.~Wampler}, \bibinfo{author}{K.~Wilson},
\newblock \bibinfo{title}{Tritium inventory in iter plasma-facing materials and tritium removal procedures},
\newblock \bibinfo{journal}{Plasma Physics and Controlled Fusion} \bibinfo{volume}{50} (\bibinfo{year}{2008}) \bibinfo{pages}{103001}.
%Type = Article
\bibitem[{Doerner et~al.(2019)Doerner, Tynan, and Schmid}]{doerner2019implications}
\bibinfo{author}{R.~Doerner}, \bibinfo{author}{G.~Tynan}, \bibinfo{author}{K.~Schmid},
\newblock \bibinfo{title}{Implications of pmi and wall material choice on fusion reactor tritium self-sufficiency},
\newblock \bibinfo{journal}{Nuclear Materials and Energy} \bibinfo{volume}{18} (\bibinfo{year}{2019}) \bibinfo{pages}{56--61}.
%Type = Article
\bibitem[{Murdoch et~al.(1999)Murdoch, Day, Gierszewski, Penzhorn, and Wu}]{murdoch1999tritium}
\bibinfo{author}{D.~Murdoch}, \bibinfo{author}{C.~Day}, \bibinfo{author}{P.~Gierszewski}, \bibinfo{author}{R.~Penzhorn}, \bibinfo{author}{C.~Wu},
\newblock \bibinfo{title}{Tritium inventory issues for future reactors: Choices, parameters, limits},
\newblock \bibinfo{journal}{Fusion Engineering and Design} \bibinfo{volume}{46} (\bibinfo{year}{1999}) \bibinfo{pages}{255--271}.
%Type = Article
\bibitem[{Schmid et~al.(2012)Schmid, Rieger, and Manhard}]{schmid2012hydrogen}
\bibinfo{author}{K.~Schmid}, \bibinfo{author}{V.~Rieger}, \bibinfo{author}{A.~Manhard},
\newblock \bibinfo{title}{Hydrogen retention in tungsten exposed to high-flux plasma},
\newblock \bibinfo{journal}{Journal of Nuclear Materials} \bibinfo{volume}{426} (\bibinfo{year}{2012}) \bibinfo{pages}{247--253}.
%Type = Article
\bibitem[{Ogorodnikova et~al.(2008)Ogorodnikova, Roth, and Mayer}]{ogorodnikova2008ion}
\bibinfo{author}{O.~V. Ogorodnikova}, \bibinfo{author}{J.~Roth}, \bibinfo{author}{M.~Mayer},
\newblock \bibinfo{title}{Ion-driven deuterium retention in tungsten},
\newblock \bibinfo{journal}{Journal of Applied Physics} \bibinfo{volume}{103} (\bibinfo{year}{2008}) \bibinfo{pages}{034902}.
%Type = Article
\bibitem[{Frauenfelder(1969)}]{frauenfelder1969solution}
\bibinfo{author}{R.~Frauenfelder},
\newblock \bibinfo{title}{Solution and diffusion of hydrogen in tungsten},
\newblock \bibinfo{journal}{Journal of Vacuum Science and Technology} \bibinfo{volume}{6} (\bibinfo{year}{1969}) \bibinfo{pages}{388--397}.
%Type = Article
\bibitem[{Philipps(2011)}]{philipps2011tungsten}
\bibinfo{author}{V.~Philipps},
\newblock \bibinfo{title}{Tungsten as material for plasma-facing components in fusion devices},
\newblock \bibinfo{journal}{Journal of Nuclear Materials} \bibinfo{volume}{415} (\bibinfo{year}{2011}) \bibinfo{pages}{S2--S9}.
%Type = Article
\bibitem[{Wright et~al.(2012)Wright, Doerner, and et~al.}]{wright2012hydrogenic}
\bibinfo{author}{G.~M. Wright}, \bibinfo{author}{R.~P. Doerner}, \bibinfo{author}{et~al.},
\newblock \bibinfo{title}{Hydrogenic retention of high-z refractory metals exposed to iter divertor-relevant plasma conditions},
\newblock \bibinfo{journal}{Nuclear Fusion} \bibinfo{volume}{52} (\bibinfo{year}{2012}) \bibinfo{pages}{042003}.
%Type = Article
\bibitem[{Longhurst and Ambrosek(2005)}]{longhurst2005verification}
\bibinfo{author}{G.~R. Longhurst}, \bibinfo{author}{J.~Ambrosek},
\newblock \bibinfo{title}{Verification and validation of the tritium transport code tmap7},
\newblock \bibinfo{journal}{Fusion Science and Technology} \bibinfo{volume}{48} (\bibinfo{year}{2005}) \bibinfo{pages}{468--471}.
%Type = Article
\bibitem[{Hu et~al.(2017)Hu, Wirth, and Maroudas}]{hu2017thermal}
\bibinfo{author}{L.~Hu}, \bibinfo{author}{B.~D. Wirth}, \bibinfo{author}{D.~Maroudas},
\newblock \bibinfo{title}{Thermal conductivity of tungsten: effects of plasma-related structural defects from molecular-dynamics simulations},
\newblock \bibinfo{journal}{Applied Physics Letters} \bibinfo{volume}{111} (\bibinfo{year}{2017}) \bibinfo{pages}{081902}.
%Type = Article
\bibitem[{Wang and et~al.(2021)}]{wang2021helium}
\bibinfo{author}{Y.~Wang}, \bibinfo{author}{et~al.},
\newblock \bibinfo{title}{Effect of helium pre-implantation on the thermal shock performance of tungsten},
\newblock \bibinfo{journal}{Nuclear Materials and Energy} \bibinfo{volume}{27} (\bibinfo{year}{2021}) \bibinfo{pages}{100934}.
%Type = Article
\bibitem[{Wirtz et~al.(2016)Wirtz, Berger, Huber, Kreter, Linke, Pintsuk, Rasinski, Sergienko, and Unterberg}]{wirtz2016helium}
\bibinfo{author}{M.~Wirtz}, \bibinfo{author}{M.~Berger}, \bibinfo{author}{A.~Huber}, \bibinfo{author}{A.~Kreter}, \bibinfo{author}{J.~Linke}, \bibinfo{author}{G.~Pintsuk}, \bibinfo{author}{M.~Rasinski}, \bibinfo{author}{G.~Sergienko}, \bibinfo{author}{B.~Unterberg},
\newblock \bibinfo{title}{Influence of helium induced nanostructures on the thermal shock performance of tungsten},
\newblock \bibinfo{journal}{Nuclear Materials and Energy} \bibinfo{volume}{9} (\bibinfo{year}{2016}) \bibinfo{pages}{177--180}.
%Type = Article
\bibitem[{Longhurst(1985)}]{longhurst1985soret}
\bibinfo{author}{G.~Longhurst},
\newblock \bibinfo{title}{The soret effect and its implications for fusion reactors},
\newblock \bibinfo{journal}{Journal of Nuclear Materials} \bibinfo{volume}{131} (\bibinfo{year}{1985}) \bibinfo{pages}{61--69}.
%Type = Article
\bibitem[{Baskes et~al.(1982)Baskes, Bauer, and Wilson}]{baskes1982tritium}
\bibinfo{author}{M.~Baskes}, \bibinfo{author}{W.~Bauer}, \bibinfo{author}{K.~Wilson},
\newblock \bibinfo{title}{Tritium permeation in fusion reactor first walls},
\newblock \bibinfo{journal}{Journal of Nuclear Materials} \bibinfo{volume}{111-112} (\bibinfo{year}{1982}) \bibinfo{pages}{663--666}.
%Type = Article
\bibitem[{Sugisaki et~al.(1982)Sugisaki, Mukai, Idemitsu, and Furuya}]{sugisaki1982isotope}
\bibinfo{author}{M.~Sugisaki}, \bibinfo{author}{S.~Mukai}, \bibinfo{author}{K.~Idemitsu}, \bibinfo{author}{H.~Furuya},
\newblock \bibinfo{title}{Isotope effect in heat of transport of h, d and t in nb},
\newblock \bibinfo{journal}{Journal of Nuclear Materials}  (\bibinfo{year}{1982}). \bibinfo{note}{Department of Nuclear Engineering, Kyushu University}.
%Type = Article
\bibitem[{Markelj et~al.(2020)Markelj, Schwarz-Selinger, Pečovnik, Chrominski, Šestan, and Zavašnik}]{markelj2020deuterium}
\bibinfo{author}{S.~Markelj}, \bibinfo{author}{T.~Schwarz-Selinger}, \bibinfo{author}{M.~Pečovnik}, \bibinfo{author}{W.~Chrominski}, \bibinfo{author}{A.~Šestan}, \bibinfo{author}{J.~Zavašnik},
\newblock \bibinfo{title}{Deuterium transport and retention in the bulk of tungsten containing helium: the effect of helium concentration and microstructure},
\newblock \bibinfo{journal}{Nuclear Fusion} \bibinfo{volume}{60} (\bibinfo{year}{2020}) \bibinfo{pages}{106029}.
%Type = Article
\bibitem[{Asaro et~al.(2007)Asaro, Farkas, and Kulkarni}]{asaro2007soret}
\bibinfo{author}{R.~J. Asaro}, \bibinfo{author}{D.~Farkas}, \bibinfo{author}{Y.~Kulkarni},
\newblock \bibinfo{title}{The soret effect in diffusion in crystals},
\newblock \bibinfo{journal}{Acta Materialia}  (\bibinfo{year}{2007}). \bibinfo{note}{UC San Diego and Virginia Tech}.
%Type = Article
\bibitem[{Dasgupta et~al.(2023)Dasgupta, Blondel, Mart{\'i}nez, Maroudas, and Wirth}]{dasgupta2023soret}
\bibinfo{author}{D.~Dasgupta}, \bibinfo{author}{S.~Blondel}, \bibinfo{author}{E.~Mart{\'i}nez}, \bibinfo{author}{D.~Maroudas}, \bibinfo{author}{B.~D. Wirth},
\newblock \bibinfo{title}{Impact of soret effect on hydrogen and helium retention in pfc tungsten under elm-like conditions},
\newblock \bibinfo{journal}{Nuclear Fusion} \bibinfo{volume}{63} (\bibinfo{year}{2023}) \bibinfo{pages}{076029}.
%Type = Article
\bibitem[{Heinola and Ahlgren(2010)}]{heinola2010diffusion}
\bibinfo{author}{K.~Heinola}, \bibinfo{author}{T.~Ahlgren},
\newblock \bibinfo{title}{Diffusion of hydrogen in bcc tungsten studied with first principle calculations},
\newblock \bibinfo{journal}{Journal of Applied Physics} \bibinfo{volume}{107} (\bibinfo{year}{2010}) \bibinfo{pages}{113531}.
%Type = Article
\bibitem[{Schmid et~al.(2017)Schmid, Bauer, Schwarz-Selinger, Markelj, von Toussaint, Manhard, and Jacob}]{schmid2017recent}
\bibinfo{author}{K.~Schmid}, \bibinfo{author}{J.~Bauer}, \bibinfo{author}{T.~Schwarz-Selinger}, \bibinfo{author}{S.~Markelj}, \bibinfo{author}{U.~von Toussaint}, \bibinfo{author}{A.~Manhard}, \bibinfo{author}{W.~Jacob},
\newblock \bibinfo{title}{Recent progress in the understanding of h transport and trapping in w},
\newblock \bibinfo{journal}{Physica Scripta} \bibinfo{volume}{2017} (\bibinfo{year}{2017}) \bibinfo{pages}{014037}.
%Type = Article
\bibitem[{Daw et~al.(1993)Daw, Foiles, and Baskes}]{daw1993embedded}
\bibinfo{author}{M.~S. Daw}, \bibinfo{author}{S.~M. Foiles}, \bibinfo{author}{M.~I. Baskes},
\newblock \bibinfo{title}{The embedded-atom method: a review of theory and applications},
\newblock \bibinfo{journal}{Materials Science Reports} \bibinfo{volume}{9} (\bibinfo{year}{1993}) \bibinfo{pages}{251--310}.
%Type = Article
\bibitem[{Marinica et~al.(2013)Marinica, Ventelon, Gilbert, Proville, Dudarev, Bencteux, and Willaime}]{Marinica2013EAM}
\bibinfo{author}{M.-C. Marinica}, \bibinfo{author}{L.~Ventelon}, \bibinfo{author}{M.~Gilbert}, \bibinfo{author}{L.~Proville}, \bibinfo{author}{S.~Dudarev}, \bibinfo{author}{G.~Bencteux}, \bibinfo{author}{F.~Willaime},
\newblock \bibinfo{title}{Interatomic potentials for modelling radiation defects and dislocations in tungsten},
\newblock \bibinfo{journal}{Journal of physics. Condensed matter : an Institute of Physics journal} \bibinfo{volume}{25} (\bibinfo{year}{2013}) \bibinfo{pages}{395502}.
%Type = Article
\bibitem[{Plimpton(1995)}]{plimpton1995fast}
\bibinfo{author}{S.~Plimpton},
\newblock \bibinfo{title}{Fast parallel algorithms for short-range molecular dynamics},
\newblock \bibinfo{journal}{Journal of Computational Physics} \bibinfo{volume}{117} (\bibinfo{year}{1995}) \bibinfo{pages}{1--19}.
%Type = Article
\bibitem[{Hu et~al.(2014)Hu, Hammond, and Wirth}]{hu2014helium}
\bibinfo{author}{L.~Hu}, \bibinfo{author}{K.~D. Hammond}, \bibinfo{author}{B.~D. Wirth},
\newblock \bibinfo{title}{Interactions of mobile helium clusters with surfaces and grain boundaries of plasma-exposed tungsten},
\newblock \bibinfo{journal}{Journal of Applied Physics} \bibinfo{volume}{115} (\bibinfo{year}{2014}) \bibinfo{pages}{173512}.
%Type = Article
\bibitem[{Hu et~al.(2015)Hu, Hammond, and Wirth}]{hu2015mdhelium}
\bibinfo{author}{L.~Hu}, \bibinfo{author}{K.~D. Hammond}, \bibinfo{author}{B.~D. Wirth},
\newblock \bibinfo{title}{Molecular-dynamics analysis of mobile helium cluster reactions near surfaces of plasma-exposed tungsten},
\newblock \bibinfo{journal}{Journal of Applied Physics} \bibinfo{volume}{118} (\bibinfo{year}{2015}) \bibinfo{pages}{163301}.
%Type = Article
\bibitem[{Fernandez et~al.(2015)Fernandez, Ferro, and Kato}]{fernandez2015dftdiffusion}
\bibinfo{author}{N.~Fernandez}, \bibinfo{author}{Y.~Ferro}, \bibinfo{author}{D.~Kato},
\newblock \bibinfo{title}{Hydrogen diffusion and vacancies formation in tungsten: Density functional theory calculations and statistical models},
\newblock \bibinfo{journal}{Acta Materialia} \bibinfo{volume}{94} (\bibinfo{year}{2015}) \bibinfo{pages}{307--318}.
%Type = Article
\bibitem[{Martínez et~al.(2021)Martínez, Mathew, Perez, Blondel, Dasgupta, Wirth, and Maroudas}]{martinez2021thermal}
\bibinfo{author}{E.~Martínez}, \bibinfo{author}{N.~Mathew}, \bibinfo{author}{D.~Perez}, \bibinfo{author}{S.~Blondel}, \bibinfo{author}{D.~Dasgupta}, \bibinfo{author}{B.~Wirth}, \bibinfo{author}{D.~Maroudas},
\newblock \bibinfo{title}{Thermal gradient effect on helium and self-interstitial transport in tungsten},
\newblock \bibinfo{journal}{Journal of Applied Physics} \bibinfo{volume}{130} (\bibinfo{year}{2021}) \bibinfo{pages}{215904}.
%Type = Article
\bibitem[{Virtanen et~al.(2020)Virtanen, Gommers, Oliphant, Haberland, Reddy, Cournapeau, Burovski, Peterson, Weckesser, Bright, {van der Walt}, Brett, Wilson, Millman, Mayorov, Nelson, Jones, Kern, Larson, Carey, Polat, Feng, Moore, {VanderPlas}, Laxalde, Perktold, Cimrman, Henriksen, Quintero, Harris, Archibald, Ribeiro, Pedregosa, {van Mulbregt}, and {SciPy 1.0 Contributors}}]{2020SciPy-NMeth}
\bibinfo{author}{P.~Virtanen}, \bibinfo{author}{R.~Gommers}, \bibinfo{author}{T.~E. Oliphant}, \bibinfo{author}{M.~Haberland}, \bibinfo{author}{T.~Reddy}, \bibinfo{author}{D.~Cournapeau}, \bibinfo{author}{E.~Burovski}, \bibinfo{author}{P.~Peterson}, \bibinfo{author}{W.~Weckesser}, \bibinfo{author}{J.~Bright}, \bibinfo{author}{S.~J. {van der Walt}}, \bibinfo{author}{M.~Brett}, \bibinfo{author}{J.~Wilson}, \bibinfo{author}{K.~J. Millman}, \bibinfo{author}{N.~Mayorov}, \bibinfo{author}{A.~R.~J. Nelson}, \bibinfo{author}{E.~Jones}, \bibinfo{author}{R.~Kern}, \bibinfo{author}{E.~Larson}, \bibinfo{author}{C.~J. Carey}, \bibinfo{author}{{\.I}.~Polat}, \bibinfo{author}{Y.~Feng}, \bibinfo{author}{E.~W. Moore}, \bibinfo{author}{J.~{VanderPlas}}, \bibinfo{author}{D.~Laxalde}, \bibinfo{author}{J.~Perktold}, \bibinfo{author}{R.~Cimrman}, \bibinfo{author}{I.~Henriksen}, \bibinfo{author}{E.~A. Quintero}, \bibinfo{author}{C.~R. Harris}, \bibinfo{author}{A.~M. Archibald}, \bibinfo{author}{A.~H. Ribeiro},
  \bibinfo{author}{F.~Pedregosa}, \bibinfo{author}{P.~{van Mulbregt}}, \bibinfo{author}{{SciPy 1.0 Contributors}},
\newblock \bibinfo{title}{{{SciPy} 1.0: Fundamental Algorithms for Scientific Computing in Python}},
\newblock \bibinfo{journal}{Nature Methods} \bibinfo{volume}{17} (\bibinfo{year}{2020}) \bibinfo{pages}{261--272}.
%Type = Article
\bibitem[{Schottky(1963)}]{schottky1963thermal}
\bibinfo{author}{C.~Schottky},
\newblock \bibinfo{title}{A theory of thermal diffusion based on the lattice dynamics of a linear chain},
\newblock \bibinfo{journal}{physica status solidi (b)} \bibinfo{volume}{3} (\bibinfo{year}{1963}) \bibinfo{pages}{357}.

\end{thebibliography}


\end{document}